\documentclass[letterpaper, 10 pt, conference]{ieeeconf}  
\IEEEoverridecommandlockouts                              

\usepackage{graphics}    
\usepackage{times}       
\usepackage{amsmath}     
\usepackage{amssymb}     
\usepackage{graphicx}
\usepackage{tabularx}
\usepackage{booktabs}
\usepackage{multirow}

\usepackage[separate-uncertainty=true,number-unit-product=\ ]{siunitx}%
\usepackage{subcaption}
\usepackage{tabularx}
\usepackage[protrusion=true,expansion]{microtype}
\usepackage{caption}
\usepackage[hidelinks]{hyperref} 
\usepackage{makecell}
\usepackage{flushend}

\def\eqref#1{Eq.~(\ref{#1})}

\newcommand\etal{~\emph{et al.~}}

\title{\LARGE \bf The Impact of VR and 2D Interfaces on Human Feedback \\ in Preference-Based Robot Learning}

\author{Jorge de Heuvel$^{1,3,4}$ \and Daniel Marta$^2$ \and Simon Holk$^2$ \and Iolanda Leite$^2$ \and Maren Bennewitz$^{1,3,4, 5}$
\thanks{
$^1$: Humanoid Robots Lab, University of Bonn, Germany.
$^2$: Division of Robotics, Perception and Learning, KTH Royal Institute of Technology, Sweden.
$^3$: Robotics Institute Germany.
$^4$: The Lamarr Institute, Bonn, Germany.
$^5$: Center for Robotics, University of Bonn, Germany.
We are grateful to Marlene Wessels for her valuable comments on the manuscript, Tharun Sethuraman for his assistance with interface development, and Subham Agrawal for his help in proofreading the manuscript.
This work has partially been funded the Robotics Institute Germany, grant No. 16ME0999, by grants from the Swedish Research Council (2024-05867), the Swedish Foundation for Strategic Research (SSF FFL18-0199), the Digital Futures research center, the Vinnova Competence Center for Trustworthy Edge Computing Systems and Applications at KTH, and the Wallenberg Al, Autonomous Systems and Software Program (WASP) funded by the Knut and Alice Wallenberg Foundation.}
}

\begin{document}
\maketitle
\thispagestyle{empty} 
\pagestyle{empty}

\begin{abstract} 
Aligning robot navigation with human preferences is essential for ensuring comfortable, and predictable robot movement in shared spaces.
While preference-based learning methods, such as reinforcement learning from human feedback (RLHF), enable this alignment, the choice of the preference collection interface may influence the process.
Traditional 2D interfaces provide structured views but lack spatial depth, whereas immersive VR offers richer perception, potentially affecting preference articulation. 
This study systematically examines how the interface modality impacts human preference collection and navigation policy alignment. 
We introduce a novel dataset of 2,325 human preference queries collected through both VR and 2D interfaces, revealing significant differences in user experience, preference consistency, and policy outcomes.
Our findings highlight the trade-offs between immersion, perception, and preference reliability, emphasizing the importance of interface selection in preference-based robot learning. 
The dataset is available to support future research.
\end{abstract}

\section{Introduction}
\label{sec:intro}
Effective robot navigation in human-populated environments requires alignment with human preferences to ensure safety, efficiency, and user acceptance. 
Integrating these preferences into navigation policies enhances human–robot interaction by improving comfort and personalization. 
Recent advances in preference-based learning, including reinforcement learning from human feedback (RLHF)~\cite{casper_open_2023}, demonstrate the potential of human-in-the-loop methods to shape robot behavior in alignment with user expectations.
In fact, preferences have been leveraged in robot learning across various settings, including multi-task learning~\cite{hejna2023few}, collaborative tasks~\cite{zhao2023learning}, language-based tasks~\cite{gillet2024shielding, holk2024predilect}, and social navigation~\cite{wang_feedback-efficient_2022-1}.

A key challenge in preference-based robot learning is the method of preference elicitation. 
Various interfaces have been explored to facilitate this process, with conventional 2D visualizations such as first-person and bird’s-eye-view videos being widely employed. 
While these 2D interfaces provide accessible and structured representations of navigation scenarios, they lack the depth and spatial context necessary for nuanced human judgment, particularly in complex 3D environments. 
More immersive alternatives, such as virtual reality (VR), offer a richer perceptual experience, potentially improving preference collection by providing a more realistic sense of robot motion and environmental context~\cite{tsoi2022sean2, baker_towards_2020}. 

Despite the potential advantages of VR, systematic comparisons between immersive and traditional 2D video-based interfaces for preference collection remain sparse.
Previous studies have primarily focused on individual interface performance or user engagement without thoroughly investigating how the chosen interface modality affects preference data quality and the subsequent alignment of robot navigation policies. 
A critical question is whether user preferences differ between VR and 2D interfaces, and how these differences influence navigation policies.

\begin{figure}[t]
	\centering
	\includegraphics[width=0.95\linewidth]{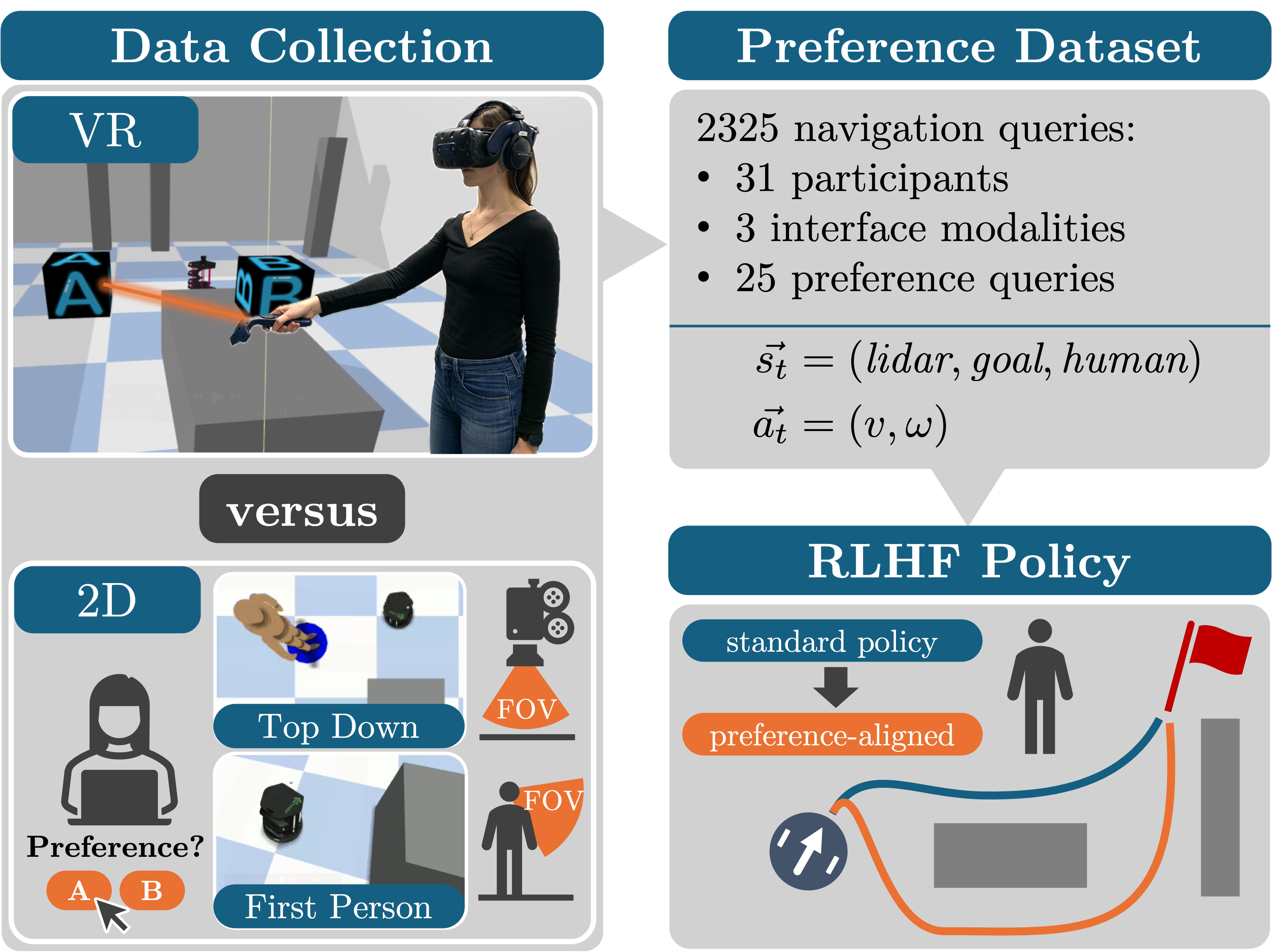}
	\caption{
	  \textbf{Left:} Our study collects preferences via Virtual Reality (VR) and 2D interfaces (top-down \& first-person views), enabling a systematic comparison of interface modalities for collecting human preferences on robot navigation.
      \textbf{Top-right:} The preference dataset consists of 2,325 navigation queries from 31 participants.
      \textbf{Bottom-right:} Using Reinforcement Learning from Human Feedback (RLHF) with our dataset, we refine a standard navigation policy (blue) into preference-aligned policies (orange).
        \label{fig:motivation}
        }
\end{figure}

In this paper, we introduce a novel user preference dataset for robot navigation, collected through both VR and 2D interfaces, see Fig.~\ref{fig:motivation}.
Our study systematically evaluates the impact of interface modality on the user experience and preference collection process.
We compare preferences elicited using a VR interface against those obtained via first-person and bird’s-eye/top-down view video interfaces. 
Finally, we derive preference-aligned navigation policies from the collected preference data, taking into account the interface modality used for collection.
Our primary contributions are:
\begin{itemize} 
	\item The collection of a dataset capturing user preferences on robot navigation across VR and 2D interfaces.
    \item A quantitative and qualitative evaluation of how different interfaces influence the collection of user preferences.
    \item The derivation and analysis of three preference-aligned navigation policies from modality-specific preferences.
\end{itemize}

 \vspace*{-0.5em}
\section{Related Work}
\label{sec:related}
\subsection{Preference Learning in Robotics}
Integrating human preferences into robotic systems has gained significant attention in recent years, particularly in human-robot interaction and autonomous navigation. 
Preference-based reinforcement learning (PbRL)~\cite{wirth2017survey, christiano2017deep, ibarz2018reward, ziegler2019fine} is a key approach that enables robots to align their behaviors with human expectations by iteratively refining a policy~\cite{amin2017repeated} through preference feedback rather than explicit reward functions. 
Wang \etal \cite{wang_personalization_2024} proposed a preference-based action representation learning (PbARL) approach that efficiently fine-tunes pre-trained policies to human preferences, allowing for effective personalization in robot behavior without requiring extensive retraining. 
Similarly, Palan \etal \cite{palan_learning_2019} introduced a hybrid learning framework that combines expert demonstrations with preference queries to improve the efficiency of reward function learning, mitigating the limitations of inverse reinforcement learning and standard PbRL methods. 
The effectiveness of learning is also contingent on the quality of queries presented to humans. 
To enhance query informativeness, various PbRL active learning techniques have been developed, leveraging policy ensembles~\cite{de2024enquery} or unsupervised learning~\cite{marta2023variquery,marta2024sequel}.

Beyond reinforcement learning, Bacchin\etal~\cite{bacchin_preference-based_2024} introduced a people-aware navigation system for telepresence robots that fuses remote operator commands with a probabilistic model of human-robot interaction. 
Their system dynamically adjusts the robot’s behavior based on inferred social signals, demonstrating how preference-aware navigation can enhance both user satisfaction and social compatibility. 
In another study, Zhou\etal\cite{zhou_exploring_2024-1} explored how human preferences can guide the improvement of inappropriate robot behaviors. 
Their findings highlight the importance of capturing nuanced human feedback to refine robot motion strategies for social navigation.

Recent works have also examined the role of virtual environments in preference learning. 
De Heuvel\etal\cite{de_heuvel_learning_2023-1} proposed a VR-based demonstration interface for learning personalized robot navigation policies, emphasizing the benefits of VR in capturing human motion preferences for dynamic environments. 
Their study highlights the utility of immersive settings in enabling intuitive and expressive demonstrations by users.

\vspace*{-0.25em}
\subsection{Interfaces in HRI}
The design of effective query interfaces for human-in-the-loop preference learning plays a crucial role in shaping the quality and reliability of collected preference data. 
Traditional preference elicitation methods often rely on 2D graphical user interfaces (GUIs), such as first-person and bird’s-eye view perspectives, which have been widely used for interactive reinforcement learning~\cite{marta2021human} and human-robot collaboration~\cite{sarah2022,marta2023aligning}. 
However, recent advancements in immersive technologies, including VR and mixed reality, have introduced novel interaction paradigms that promise more natural and context-rich preference acquisition~\cite{tsoi2022sean2}.

Wonsick and Padır \cite{wonsick_systematic_2020} classify VR interfaces for robot control into five areas: visualization, control, interaction, usability, and infrastructure. 
Their comparison of VR and traditional keyboard-mouse-monitor setups (KBM) for humanoid teleoperation shows that VR enhances engagement, intuitive control, and spatial awareness while reducing cognitive load, whereas KBM benefits from widely available hardware.

LeMasurier\etal\cite{lemasurier_comparing_2024} further investigated the trade-offs between 2D and VR interfaces for human-in-the-loop robot planning in navigation and manipulation tasks. 
Their study finds that while KBM interfaces yield higher task performance, VR interfaces lead to fewer collisions, making them preferable for high-risk scenarios where safety is paramount. 

Wozniak et al. \cite{wozniak_happily_2023} explored the effectiveness of VR interfaces for correcting robot perception errors, comparing them with traditional screen-based interfaces. 
Their study found the VR interface to be more immersive and enjoyable for the users, who preferred it over the screen-based alternative.

These studies underscore the growing relevance of immersive interfaces for robotics.
Our study therefore investigates the differences between VR and 2D KBM interfaces for user preference acquisition.

\vspace*{-0.5em}
\section{Method}
\vspace*{-0.25em}
We subsequently provide an overview on the robot's navigation task, the interface, and user study for data collection.

\vspace*{-0.25em}
\subsection{Problem Statement}
This work investigates how different user interfaces influence the user preference collection for learning-based robot behavior adaptation. 
We focus on a query interface where users provide pairwise comparisons of pre-recorded robot navigation trajectories, a key component of preference-based reinforcement learning (PbRL).
We analyze the impact of interface modalities (VR vs. 2D GUI) and scene perspective on user experience and preference expression. 
As an application scenario, we consider a human-aware robot navigation task in which a robot navigates to a goal in an environment with static obstacles and a nearby human. 
The human may have specific preferences regarding the robot's navigation behavior.

\vspace*{-0.25em}
\subsection{Learning Robot Navigation}
In line with our target methodology, PbRL, the navigation task is solved via reinforcement learning, where an agent learns to navigate the robot to the goal using velocity commands.
As a simulation environment, iGibson~\cite{li_igibson_2022} with its PyBullet physics engine is used to simulate a Kobuki Turtlebot 2i. 
The navigation scene resembles an open space with position-randomized small and large box obstacles and a static human.
Start, goal and human position are sampled in  in close proximity to each other.
The core navigation reward contains a sparse goal reaching reward, a continuous time penalty, sparse collision and timeout penalties, and penalties for self-intersecting trajectories and jerk:
\begin{equation}
r^t = r^t_\textit{goal} + r^t_\textit{time} + r^t_\textit{collision} + r^t_\textit{timeout} + r^t_\textit{loop} + r^t_\textit{jerk}
\label{eq:r_core}
\end{equation}
Episodes end upon reaching the goal, a collision, or a timeout.

\vspace*{-0.25em}
\subsection{Query Generation}
To generate queries of a navigating robot for user evaluation, we use EnQuery~\cite{de2024enquery}, an ensemble-based query generation method designed to improve the efficiency and reliability of user preference collection in PbRL. 
EnQuery is particularly suited for applications where behavior diversity is required under consistent environmental conditions, such as in a given navigation scenario.

Following \cite{de2024enquery}, we employ an ensemble $\mathcal{E} = {\pi_i(s_t, a_t) \mid i \in [N_E]}$ of $N_E = 4$ policies, referred to as the policy ensemble of TD3~\cite{fujimoto2018addressing} reinforcement learning (RL) policies. These policies are trained with a regularization term that promotes behavioral diversity among ensemble members. 
Thus, we obtain two distinct 2D trajectories connecting the same start and goal.
This approach ensures that all generated trajectory options are grounded in a common reference frame, which aims to improve re-test reliability by reducing variations in extraneous environmental factors. 

Once trained, the ensemble $\mathcal{E}$ is used to generate diverse trajectory options for the randomized scene configuration, by sampling two individual ensemble policies and rolling them out in the sampled navigation scenario.
To ensure meaningful queries, colliding or self-intersecting trajectories filtered.
We generate a dataset $D_Q$ of $N=500$ queries, from which we sample subsets for the participants to rate in the user study, either in VR or as 2D video playback.

\vspace*{-0.25em}
\subsection{Query Interfaces}
To collect user preferences, we employ three types of query interfaces: an immersive virtual reality (VR) setup and two 2D video-based KBM interfaces on a desktop computer.
Both interfaces show the same navigation environment but differ in perspective and immersion.
The VR interface provides an interactive and immersive experience, whereas the 2D video KBM interface offers a more conventional, screen-based alternative.
Within the 2D~interface, we present two perspectives: Top-down (\mbox{2D-TD}), as in \cite{marta2023aligning}, also known as a bird’s-eye view, and first-person view (\mbox{2D-FPV}), which more closely resembles a VR perspective.
For a given query, the human position is defined, serving additionally as the observer position in VR and \mbox{2D-FPV}.

\subsubsection{Virtual Reality}
The VR interface is based on PyBullet VR \cite{coumans_pybullet_2016}, ensuring native compatibility with the EnQuery training environments.
We connect an HTC Vive Pro Eye setup as VR hardware.
A transparent blue  cylinder indicates the static human observer position on the floor.
Additionally, a floating dialogue in front of the user conveys instructions and announces the upcoming trajectory with labels (A or B) for $\SI{2}{\second}$.
For preference selection, participants interact with a floating selection menu by pointing and clicking on one of two labeled boxes, A or B.  

\subsubsection{2D Video}
In contrast, the 2D video KBM interface is implemented using the open-source library Pygame \cite{noauthor_pygamepygame_2025} on a desktop computer.
The full-screen interface contains a video frame with clickable buttons positioned to the side for starting queries and selecting preferences.
To maintain consistency in visualization, all query videos are recorded from two perspectives (\mbox{2D-TD} and \mbox{2D-FPV}) at a resolution of \mbox{$720 \times 404$} pixels.

Each video begins with a $\SI{2}{\second}$ trajectory label (A or B), ensuring clear differentiation between the two options.
While the \mbox{2D-FPV} video is recorded with a $\SI{60}{\degree}$ vertical field of view that tracks the robot from the perspective of the human, the \mbox{2D-TD} perspective is set to capture the entire navigation path.
Additionally, in \mbox{2D-TD}, the human is represented by a neutral wooden mannequin 3D model.

\subsubsection{Trial}
The robot is initialized at the query-specific start position for both trajectory options.
Once the participant initiates the trial by clicking a button, the pre-recorded trajectory or video plays, and the robot navigates through the scene.
The start and goal positions are not visualized.
Each query consists of two trajectories presented sequentially.
To prevent bias or premature selection, the preference selection menu is disabled until both trajectories have been displayed.
Then, the user can select their preferred trajectory.
Queries cannot be repeated, and the user must make a selection before proceeding to the next query.

\vspace*{-0.25em}
\subsection{User Study}
We conducted a user study to compile a dataset of participants’ navigation preferences and assess user experience across three preference interfaces (VR, \mbox{2D-TD}, \mbox{2D-FPV}).
Data collection was divided into three distinct stages:
S1: Collecting preferences through navigation queries for each interface modality,
S2: Post-interface interaction questionnaires assessing user experience, and
S3: A final ranking survey comparing the three interfaces.
Before testing, all participants received detailed study information, provided written consent, and completed a demographic questionnaire.
The study was structured into three blocks in randomized order, each corresponding to one of the interfaces as an experimental condition.
Each block presented the same 25 preference queries in random order (S1), initially sampled for each participant from the query dataset $D_Q$.
By presenting identical queries across different interfaces, we could later investigate the impact of the interface modality on participants’ navigation preferences.
After each interface block, participants completed a questionnaire (S2) regarding their experience with the interface and the queries, as shown in Fig. \ref{fig:survey}.
Upon completion of all three blocks, the final ranking survey (S3), based on the Technology Acceptance Model (TAM)~\cite{davis_perceived_1989}, asked participants to rank the interfaces by perceived usefulness, intention to use, and ease of use, as shown in \ref{tab:survey_questions}.
All collected data were anonymized using a coding table for participant IDs.
Each session lasted approximately one hour.

\vspace*{-0.25em}
\subsection{Participants}
A total of 32 individuals (10 women, 22 men) participated in the study in exchange for a EUR 15 monetary compensation.
All participants reported having corrected-to-normal vision.
One participant was removed due to technical issues during data collection, leaving $N=31$ participants (10 women, 21 men).
The mean age of the sample was 24.6 years (\textit{SD} = 3.7).
Participants rated their experience with AR/VR on a 7-point Likert scale, with a mean rating of 3.1 (\textit{SD} = 1.5).
Participants also rated their experience with robotics on the same scale, yielding a mean score of 3.6 (\textit{SD} = 2.0).
The study adhered to the principles outlined in the Helsinki Declaration.

\vspace*{-0.25em}
\section{Experimental Evaluation}
\vspace*{-0.25em}
\label{sec:exp}

We propose the following hypotheses:
\textbf{H1:} The user experience differs between the interface modalities.
\textbf{H2:} The user preferences for robot navigation differ between the interface modalities.
\textbf{H3:} A preference discrepancy between interfaces reflects in the navigation behavior of interface-specific preference-aligned policies.

\vspace*{-0.25em}
\subsection{Interface Questionnaire}
\begin{figure}[t]
	\centering
	\includegraphics[width=0.97\linewidth]{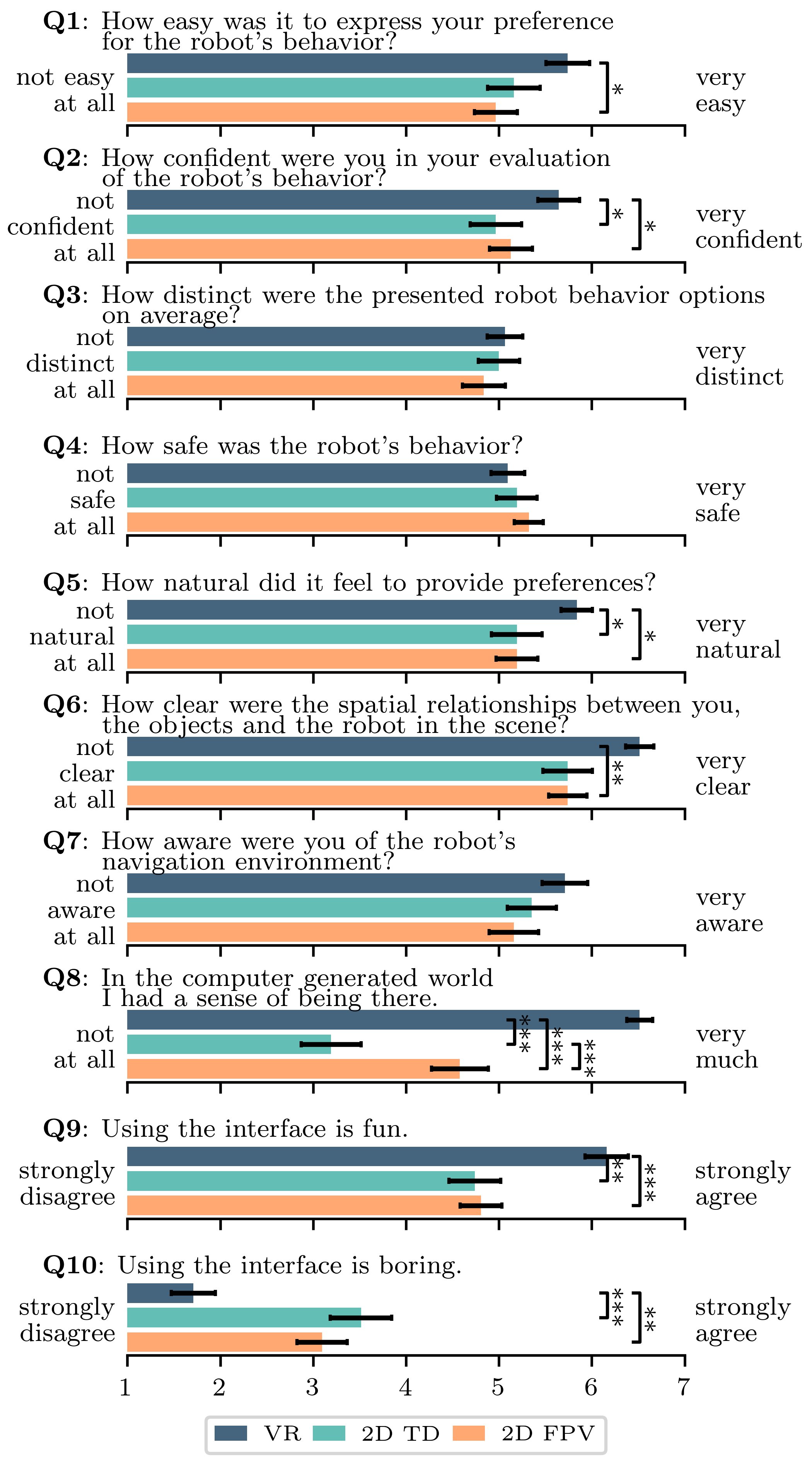}
	\caption{
	  Survey results (S2) comparing user experiences across three interface conditions: 
      virtual reality (VR), 2D top-down (\mbox{2D-TD}), and 2D first-person view (\mbox{2D-FPV}). 
    Participants rated their experience across multiple aspects after each block.
    Ratings were provided on a Likert scale (1-7), bars indicate score means, standard errors are indicated.
      Asterisks denote significance levels (* $p < .05$, ** $p < .01$, ***  $p < .001$).
        \label{fig:survey}
        }
\end{figure}
Targeting the users' interface experience and the expression of preferences, we analyze the 10-item questionnaire (Likert scale, score 1-7) of S2, see \textbf{Fig. \ref{fig:survey}}.
We used a Friedman test to statistically evaluate whether the interface modality (VR, \mbox{2D-TD}, \mbox{2D-FPV}) had a significant impact on the ratings (H1), in each of the 10 questions. 
Note that we chose a non-parametric alternative to the repeated-measures ANOVA to account for the ordinal scale level of the responses (7-point Likert scale).
We followed up with three pairwise Wilcoxon signed-rank comparisons with Bonferroni correction when the Friedman test revealed a significant impact of the interface modality.
Supporting H1, statistically significant differences in favor of the VR interface were found for the ease of expressing preferences compared to \mbox{2D-FPV} (Q1), participants'~confidence in evaluation compared to both 2D interfaces (Q2), the naturalness of providing preferences (Q5), a clearer spatial understanding compared to \mbox{2D-FPV} (Q6).
We included Q8 from a validated presence scale \cite{usoh_using_2000}, confirming higher immersion levels in VR compared to both 2D interfaces.
Participants reported that the VR interface was significantly more fun to use and less boring (Q9, Q10, System Usability Scale \cite{brooke_sus_1996}) compared to the 2D interfaces, which further supports H1.
No significant effects were observed for the remaining questions.

\vspace*{-0.25em}
\subsection{Interface Ranking}
\begin{figure}[t]
	\centering
	\includegraphics[width=1\linewidth]{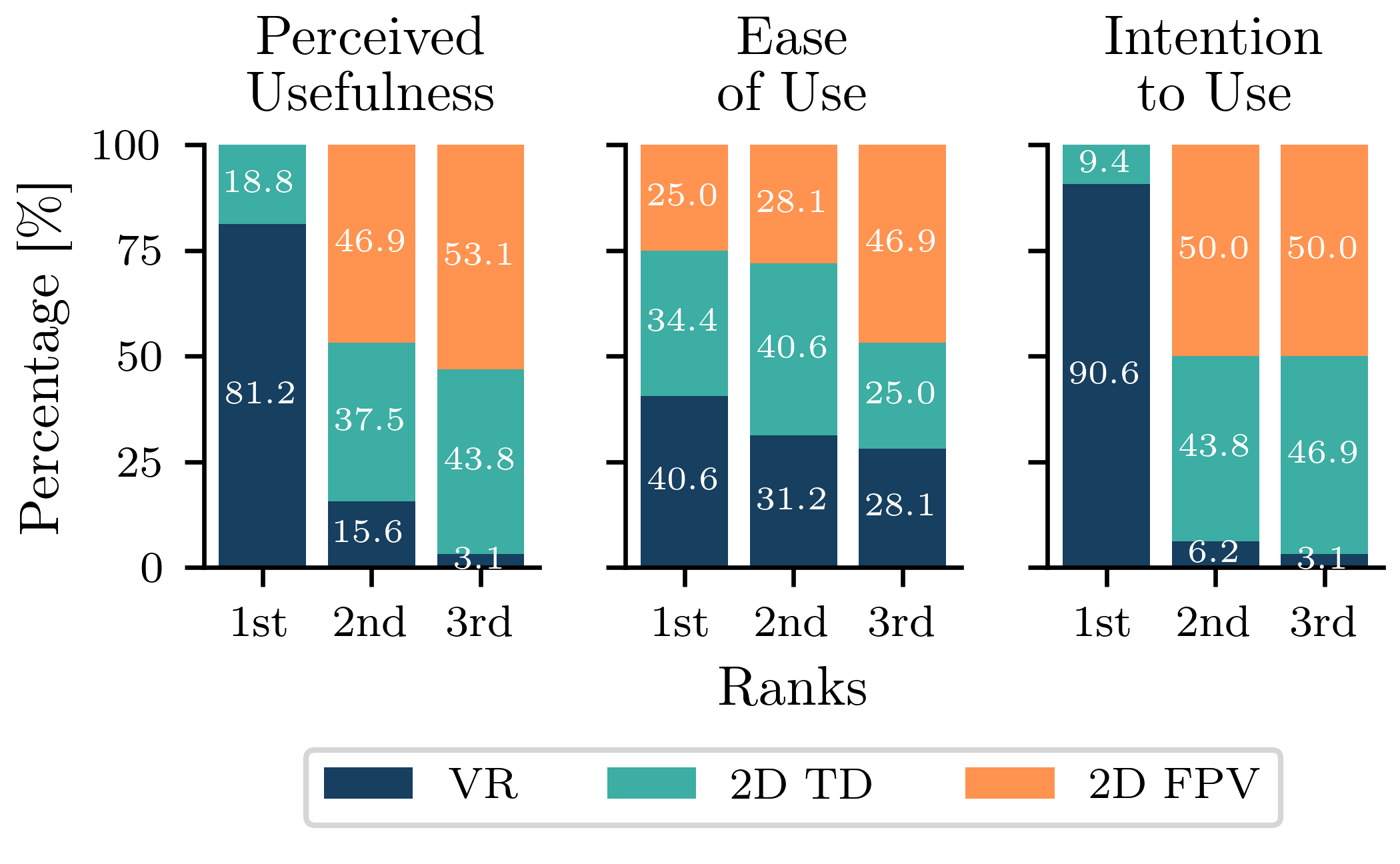}
	\caption{
	  User rankings (S3) of three modalities, namely Virtual Reality (VR), 2D Top-Down (\mbox{2D-TD}), and 2D First-Person View (\mbox{2D-FPV}), based on perceived usefulness, ease of use, and intention to use.
      Each bar represents the percentage of participants who assigned first, second, and third ranks to each modality. 
      VR is predominantly ranked highest in usefulness and intention to use.
      \label{fig:ranking}
            }
\end{figure}
After the three preference query blocks, participants ranked the interfaces (S3) based on the TAM (see \textbf{Fig. \ref{fig:ranking}} and \textbf{Table \ref{tab:survey_questions}}). 
The forced-choice ranking did not allow ties.
A chi-square test assessed deviations from equal choice distributions across VR, \mbox{2D-TD}, and \mbox{2D-FPV}. 
Significant effects were further examined using pairwise $z$-tests with Bonferroni correction.

For \textbf{Usefulness}, rankings were not evenly distributed ($p < .001$), with VR being more often preferred over both alternatives, and \mbox{2D-TD} preferred more often over \mbox{2D-FPV}. 
VR was ranked first by \SI{81.2}{\percent}, while \mbox{2D-FPV} was consistently last.

For \textbf{Ease of Use}, no significant differences were found, with rankings evenly distributed across interfaces.

For \textbf{Intention to Use}, VR was preferred more often ($p < .001$), significantly outranking both alternatives, while \mbox{2D-TD} was preferred more often over \mbox{2D-FPV}. 
Notably, \SI{90.6}{\percent} ranked VR first, and \mbox{2D-FPV} received no first-choice votes.

Overall, VR was significantly more often preferred for usefulness and intention to use, aligning with \cite{wozniak_happily_2023}, likely due to enhanced perception of robot behavior. 
Increased fun (Q9, \textbf{Fig.~\ref{fig:survey}}) may explain why the majority of participants expressed their intention to use VR. 
No interface stood out for ease of use, possibly because the preference task itself was similar to operate between interfaces. 
Across all criteria, \mbox{2D-FPV} was the least often preferred.

\begin{table}[b]
\centering
\renewcommand{\arraystretch}{1.1}
\begin{tabularx}{\linewidth}{cX}
 & \textbf{Ranking Instruction (Options: 1st, 2nd, 3rd)} \\
\hline
R1 & Order the three interfaces for their usefulness for rating the robot behavior. \\
\hline
R2 & Order the three interfaces for their ease of use. \\
\hline
R3 & Order the three interfaces based on your intention to use.\\
\hline
\end{tabularx}
\caption{Ranking instructions (S3) for the interfaces upon completion of all three interface modality blocks.}
\label{tab:survey_questions}
\end{table}

\vspace*{-0.25em}
\subsection{Dataset Overview}
The dataset is publicly available\footnote{\label{fn:dataset_url}\url{https://github.com/HumanoidsBonn/rlhf_prefnav_interface_study}} and contains 2,325 user preference queries for robot navigation, collected across three interface modalities: VR, \mbox{2D-TD}, and \mbox{2D-FPV}.

The dataset is structured by participant ID, interface modality, query, and preference.
Each query stores both trajectories and the scene configuration (e.g., for replay in VR), the 2D videos, and RL episodes (state, action, next state, reward) for both trajectories, where states  $\mathbf{s}_t = (\text{lidar}, \text{goal}, \text{human})$ capture lidar data, robot-centric goal and human position, while actions  $\mathbf{a}_t = (v, \omega)$ represent velocity commands.
Preferences are stored as A/B labels.
In addition to the state-action pairs, we provide the robot trajectory as a 2D path, the static human pose, and the obstacle poses in world coordinates.

As we later show, our dataset enables the distillation of preference models for policy alignment.

\vspace*{-0.25em}
\subsection{User Preferences}
This section deals with a quantitative analysis of the preference dataset collected in S1 with respect to the effect of the interface modality.

\subsubsection{Modality Agreement}
\begin{table}[t]
    \centering
    \begin{tabular}{ccc}
        \textbf{Modality Pair} & \textbf{Agreement [\%]} & \textbf{Standard Deviation [\%]} \\
        \hline
        VR - \mbox{2D-TD}   & 69.2 &  9.6\\  
        VR - \mbox{2D-FPV}  & 68.6 &  11.8\\  
        \mbox{2D-TD} - \mbox{2D-FPV} & 67.0 & 14.2\\  
        \hline
    \end{tabular}
    \caption{
    Mean agreement and standard deviation between different interface modalities, aggregated on a per-participant basis.
    Note that the block order of interfaces has been randomized among participants. 
    Preference changes between interfaces occur but not significantly more often for specific interface combinations.
    }
    \label{tab:modality_agreement}
\end{table}
Querying the same 25 trajectory pairs in all three interface modalities to each participant, allows us to examine whether participants exhibit different preferences between the interfaces.
We compute the interface modality agreement by matching the block-randomized queries between modalities and checking for preference agreement, see \textbf{Table~\ref{tab:modality_agreement}}.
The agreement is aggregated on a per-participant basis and subsequently averaged over all participants.
With the values of all three combinations averaging around $\SI{70}{\percent}$, we can conclude that preferences do change between interfaces, but not noticeably more often between specific interface combinations.
We conclude that consistency in the interface is key during dataset collection, as preferences can be inconsistent with an interface change. 
As reflected by the standard deviation, we observe considerable variation in interface agreement among participants.
This finding underscores the necessity for interface consistency when preference data is collected or merged.

\subsubsection{Disagreement Analysis}
\begin{figure}[t]
	\centering
	\includegraphics[width=.95\linewidth]{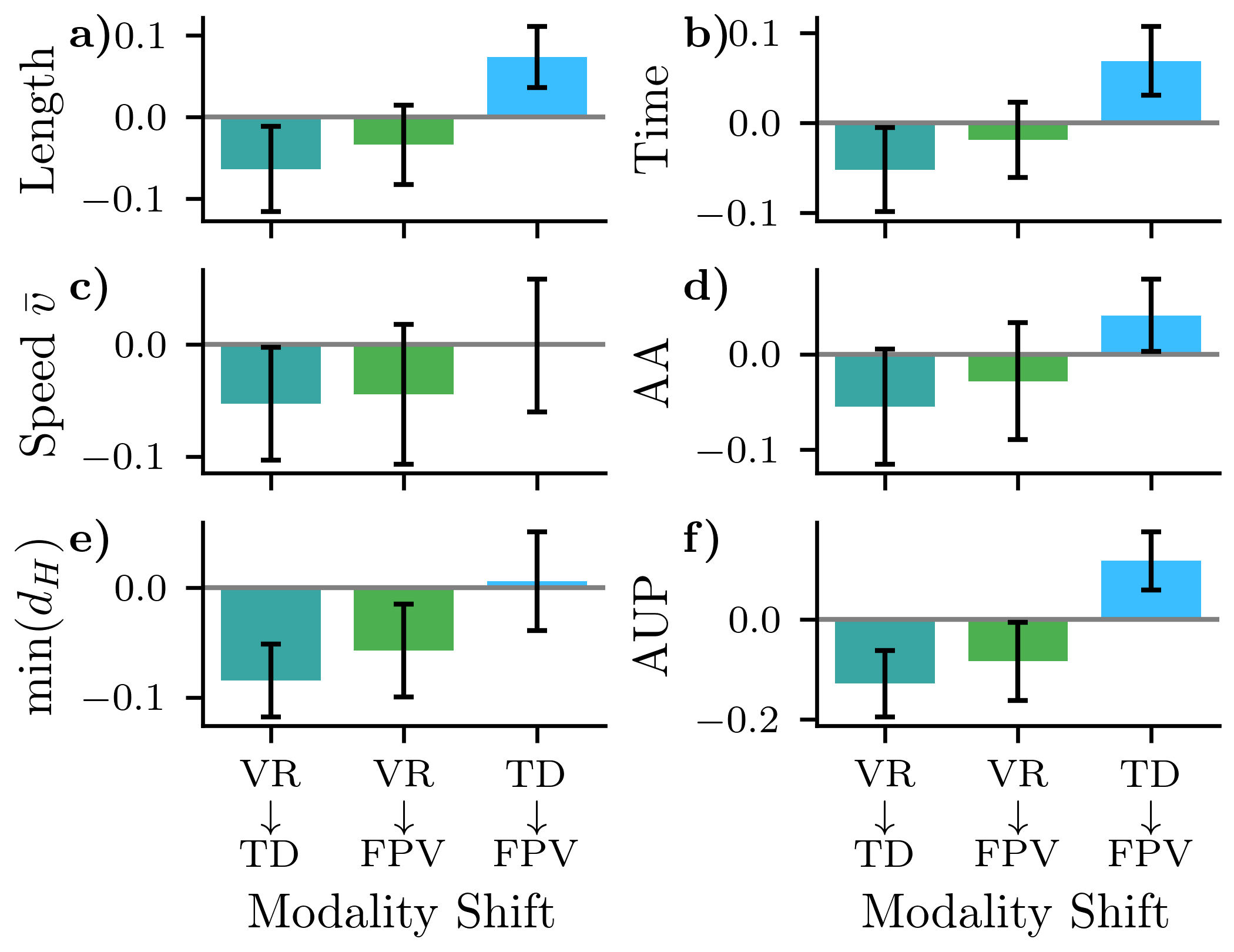}
	\caption{
	  Change in the preferred trajectory with a modality shift in cases of preference disagreement between interfaces for a given participant. 
      Metrics are $z$-standardized for all queried trajectories per participant.
      Bars show mean change, and error bars indicate the standard error of the participant means averaged over their disagreements. 
      Participants preferred shorter and more straightforward driving trajectories in 2D interfaces compared to VR, with the robot occasionally traversing closer to the human.
        \label{fig:disagreement_change}
        }
\end{figure}
We now transition from inter-modality agreement to the cases of disagreement.
When participants preferred trajectory A in one modality but trajectory B for the same AB-query in a different modality, we term this inter-modality disagreement.
For each interface combination of these inter-modality disagreements, we explore the differences between the two preferred trajectories.
Note that we report these differences but refrain from conducting inferential statistical tests because the characteristics of the two trajectories of the same query have not been experimentally controlled. 
\textbf{Fig. \ref{fig:disagreement_change}} shows the differences in preference for selected trajectory metrics, while the x-axis indicates the interface transition.
Because different participants had their own sampled subset of 25 queries with a different trajectory profile distributions, we first apply z-score normalization based on all trajectories (preferred and rejected) shown to a participant. 
Subsequently, the average differences were aggregated on a per-participant basis to account for the varying number of disagreements per participant.
We measure for changes in the trajectory length~(a), time~(b), average speed~(c), curvature/accumulated angle~(AA)~(d), minimum distance to the human (min($d_H$))~(e), and area under the path~(AUP)~(f).
The sign of the metric change corresponds to the interface transition  $A \to B$. 
For readability, only one direction is shown. The reverse transition has the same magnitude with an inverted sign.

In cases where participants exhibited different preferences for the same queries in VR and on 2D interfaces, participants preferred shorter and more straightforward driving trajectories (measured by the area under the path) in 2D compared to VR.
Additionally, the robot may traverse closer to the human, compare~(e).
Similarly, query disagreements point to a preference for more straightforward driving styles when transitioning from the \mbox{2D-FPV} to the \mbox{2D-TD} interface, see~(f).
Consequently, we find empirical support for H2.

\vspace*{-0.25em}
\subsection{Policy Alignment}
\begin{figure}[t]
	\centering
	\includegraphics[width=0.95\linewidth]{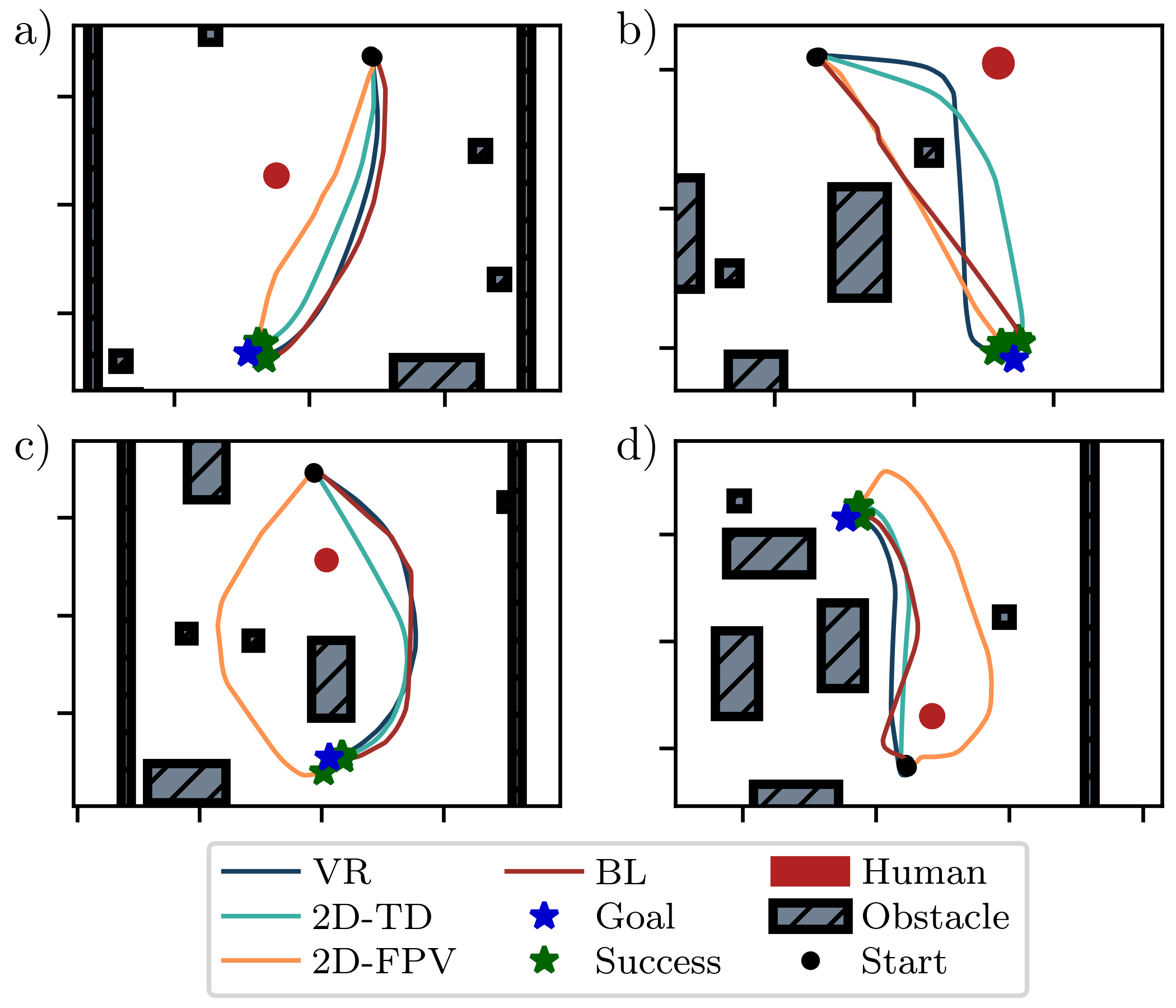}
	\caption{
	  Navigation behavior comparison between aligned policies $\pi_\text{VR}$, $\pi_\text{\mbox{2D-TD}}$, and $\pi_\text{\mbox{2D-FPV}}$ and a non-aligned baseline counterpart $\pi_\text{BL}$ in four navigation scenarios.
      The aligned policies exhibit smoother and more obstacle-aware trajectories than the non-aligned policy $\pi_\text{BL}$, with $\pi_{\text{VR}}$ and $\pi_{\text{\mbox{2D-TD}}}$ demonstrating the best balance between efficiency and safety.
        \label{fig:rlhf_policy_comparison_qualitative}
        }
\end{figure}
\begin{table}[b]
\centering
\renewcommand{\arraystretch}{1.1}
\begin{tabular}{l c c c c r}
\toprule
Metric & VR & \mbox{2D-TD} & \mbox{2D-FPV} & BL & sign. pairs\\
\midrule
CR [\%] & 5.8 & 5.8 & 4.9 & 5.2 & \\
SR [\%] & 94.2 & 94.1 & \textbf{94.9} & 94.8 & \\
TR [\%] & 0.0 & 0.1 & 0.2 & 0.0 & \\
\midrule
Steps & 49.1 & 49.5 & 55.9 & 50.0 & \texttt{\textbf{~bcd}}\\
Path Length [m] & 4.9 & 4.9 & \textbf{5.4} & 4.9 & \texttt{\textbf{~bcd}}\\
Time [s] & \textbf{9.6} & 9.7 & 11.0 & 9.8 & \texttt{\textbf{~bcd}}\\
AA [rad] & 3.2 & 3.1 & \textbf{4.1} & 2.8 & \texttt{\textbf{~bcd}}\\
AUP [m²] & 2.2 & 1.8 & \textbf{3.4} & 2.1 & \texttt{\textbf{abcd}}\\
$\min(d_h)$ [m] & \textbf{1.33} & 1.20 & 1.28 & 1.25 & \texttt{\textbf{a~c~}}\hspace{1em}\\
Speed $\bar{v}$ [m/s] & 0.51 & 0.50 & 0.49 & 0.51 & \texttt{\textbf{~bcd}}\\
\bottomrule
\end{tabular}
\caption{
Quantitative metrics for our preference-aligned $\pi_\text{VR}$, $\pi_\text{\mbox{2D-TD}}$, and $\pi_\text{\mbox{2D-FPV}}$ and a non-aligned baseline $\pi_\text{BL}$, averaged over 1,000 trajectories in randomly sampled scene configurations.
Significance pair abbreviations ($p<.05$) for unequal mean (Welch t-test, Bonf. corr.) are
\texttt{\textbf{a}}: VR vs. 2D-TD,
\texttt{\textbf{b}}: VR vs. 2D-FPV,
\texttt{\textbf{c}}: 2D-TD vs. 2D-FPV,
\texttt{\textbf{d}}: 2D-FPV vs. BL.
}
\label{tab:policy_metrics}
\end{table}
To examine how preferences collected from different interfaces impact policy alignment and navigation behavior, we employ a preference-based reinforcement learning (PbRL) approach. 
This method trains reward functions from the collected preference dataset, enabling the subsequent alignment of human-aware RL navigation policies for each query interface.
Following \cite{christiano2017deep,wirth2017survey}, we learn a parametric reward function \(\hat{r}_\psi\) from human preferences. 
For policy optimization, we implement a PbRL algorithm following \cite{christiano2017deep}, using TD3~\cite{fujimoto2018addressing} as the base RL algorithm. 
Splitting up our preference dataset by interface, we train three individual reward functions modeled as an MLP with $[256,256,256]$ hidden units based on the queries collected from the participants. 
The policies for each condition were then trained for 500k time steps by weighting the original navigation task reward and their respective preference reward model as $r = \lambda \hat{r}_\psi + (1-\lambda) r_\text{core}$, with $\lambda=0.2$.
The value of $\lambda$ was empirically determined through a grid search to ensure optimal training stability.
Note that the preference reward ranges from -1 to +1, while the original task reward from of -10 to +20, with these extrema driven by the sparse terms of $r_\text{core}$.
In addition to the trained policies $\pi_\text{VR}$, $\pi_\text{\mbox{2D-TD}}$, and $\pi_\text{\mbox{2D-FPV}}$, we also include a non-aligned baseline policy $\pi_\text{BL}$, trained on the same task using $r_\text{core}$ without preference-based rewards, for comparison.

\textbf{Fig.~\ref{fig:rlhf_policy_comparison_qualitative}} illustrates navigation trajectories of the policies in four distinct scenarios.
The plots illustrate the paths taken by each policy from the start to goal, navigating around static obstacles and avoiding the human.
All policies successfully navigate the scenes, while the aligned policies exhibit smoother and more obstacle-aware trajectories compared to their non-aligned counterpart $\pi_\text{BL}$.
In Scenario~d), both $\pi_\text{BL}$ and $\pi_\text{\mbox{2D-FPV}}$ are prone to inefficient routes. 
$\pi_\text{\mbox{2D-FPV}}$ shows the same conservative behavior in Scenario~c) as well.
Overall, the results indicate that the aligned policies $\pi_{\text{VR}}$ and $\pi_{\text{\mbox{2D-TD}}}$ achieve a superior trade-off between efficiency and safety.

The quantitative results in \textbf{Table~\ref{tab:policy_metrics}} confirm that preference-aligned policies improve human-aware navigation, with $\pi_{\text{VR}}$ achieving the best balance between efficiency and safety by maintaining the highest human clearance (1.33 m) while ensuring low travel times.
$\pi_{\text{\mbox{2D-FPV}}}$ prioritizes safety, exhibiting the lowest collision rate (4.9\%) but at the cost of longer paths (5.4 m) and higher angular accumulation (4.1 rad). 
$\pi_{\text{\mbox{2D-TD}}}$ exhibits the most straight-forward navigation, reflected in the lowest area under path~(AUP).
These findings, supported by the qualitative analysis, highlight that $\pi_{\text{VR}}$ offers the most balanced navigation performance, while $\pi_{\text{\mbox{2D-FPV}}}$ navigates overly conservatively, and $\pi_{\text{BL}}$ is less cautious.
Statistical significance on the pairwise differences between policies are marked in Table~\ref{tab:policy_metrics} in column ``sign. pairs'' with indicators \texttt{\textbf{abcd}} based on Welch’s $t$-tests with Bonferroni correction.
In summary, we find support for H3.

\vspace*{-0.25em}
\section{Conclusion}
\vspace*{-0.25em}
\label{sec:conclusion}
Our study systematically examined how interface modality affects human preference collection for robot navigation and the resulting policy alignment.
The results confirm that the choice of interface modality significantly impacts how users express preferences, perceive the interaction, and ultimately shape robot navigation behavior. 
The VR interface provided a more immersive and intuitive experience, leading to greater confidence and ease in preference expression. 
This aligns with prior findings on the benefits of immersive environments for user engagement \cite{schone_experiences_2019}. 
However, the study also revealed that preferences were not entirely consistent between interfaces, with participants favoring shorter, more direct paths in 2D interfaces while exhibiting greater tolerance for curved trajectories with increased human clearance in VR. 
This suggests that the visualization and spatial representation of the robot’s movement influence user preferences, highlighting the importance of maintaining interface consistency during preference collection. 
The navigation policies trained on interface-specific preferences demonstrated noticeable differences in robot behavior.
These findings highlight the necessity of considering interface effects in preference-based reinforcement learning, as user preference shifts due to modality changes can directly impact policy training outcomes. 
To support further research on these effects and the collected user preferences, our preference dataset is available to the research community.

\vspace*{-0.5em}
\bibliographystyle{IEEEtran}
\bibliography{bib_vr_rlhf, bib_daniel}

\end{document}